\documentclass{aa}
\pdfoutput=1
\usepackage[varg]{txfonts}
\usepackage{graphics,graphicx,enumitem}
\usepackage{amsmath,amsfonts,amssymb,mathrsfs}
\usepackage{float}
\usepackage{hyperref}
\usepackage{array}
\usepackage[table,x11names]{xcolor}
\definecolor{vLightGray}{gray}{0.85}
\usepackage{makecell}
\hypersetup{
  colorlinks   = true, 
  urlcolor     = blue, 
  linkcolor    = blue, 
  citecolor   = blue 
}
\usepackage{natbib}
\usepackage{braket}
\usepackage{changepage}
\bibpunct{(}{)}{;}{a}{}{,} 
\newcommand{\lD}{\lambda/D}
\newcommand{\FT}{\mathscr{F}}

\title{Laboratory verification of `Fast \& Furious' phase diversity: Towards controlling the low wind effect in the SPHERE instrument}
\titlerunning{F\&F for LWE @ SPHERE II}

\author{M.J. Wilby \inst{\ref{Leiden}} \and C.U. Keller \inst{\ref{Leiden}} \and J.-F. Sauvage \inst{\ref{LAM}}$^,$\inst{\ref{ONERA}} \and K. Dohlen\inst{\ref{LAM}} \and T. Fusco \inst{\ref{ONERA}} \and D. Mouillet\inst{\ref{IPAG}} \and J.-L. Beuzit\inst{\ref{IPAG}} }

\institute{{Leiden Observatory, Leiden University, PO Box 9513, 2300
RA Leiden, The Netherlands}\label{Leiden}
\and {Aix Marseille Univ, CNRS, LAM, Laboratoire d'Astrophysique de Marseille, Marseille, France}\label{LAM}
\and {ONERA, 29 Avenue de la Division Leclerc, 92320 Châtillon, France}\label{ONERA}
\and {Univ. Grenoble Alpes, CNRS, IPAG, 38000 Grenoble, France}\label{IPAG} \\
\email{wilby@strw.leidenuniv.nl}}

\date{Received: September 15, 2017 \\
Accepted: March 4, 2018}
\abstract {The low wind effect (LWE) refers to a characteristic set of quasi-static wavefront aberrations seen consistently by the SPHERE instrument when dome-level wind speeds drop below $\rm 3~ms^{-1}$. The LWE produces bright low-order speckles in the stellar PSF, which severely limit the contrast performance of SPHERE under otherwise optimal observing conditions. }
{In this paper we propose the Fast \& Furious (F\&F) phase diversity algorithm as a viable software-only solution for real-time LWE compensation, which would utilise image sequences from the SPHERE differential tip-tilt sensor (DTTS) and apply corrections via reference slope offsets on the AO system's Shack-Hartmann wavefront sensor.}
{We evaluated the closed-loop performance of F\&F on the MITHIC high-contrast test-bench, under conditions emulating LWE-affected DTTS images. These results were contrasted with predictive simulations for a variety of convergence tests, in order to assess the expected performance of an on-sky implementation of F\&F in SPHERE.}
{The algorithm was found to be capable of returning LWE-affected images to Strehl ratios of greater than 90\% within five iterations, for all appropriate laboratory test cases. These results are highly representative of predictive simulations, and demonstrate stability of the algorithm against a wide range of factors including low image signal-to-noise ratio (S/N), small image field of view, and amplitude errors. It was also found in simulation that closed-loop stability can be preserved down to image S/N as low as five while still improving overall wavefront quality, allowing for reliable operation even on faint targets.}
{The Fast \& Furious algorithm is an extremely promising solution for real-time compensation of the LWE, which can operate simultaneously with science observations and may be implemented in SPHERE without requiring additional hardware. The robustness and relatively large effective dynamic range of F\&F also make it suitable for general wavefront optimisation applications, including the co-phasing of segmented ELT-class telescopes.}


\begin{document} 	
\maketitle

\section{Introduction}
\label{Sec:Intro}

	The Spectro-Polarimetric High-Contrast Exoplanet REsearch instrument (SPHERE) \citep{Beuzit:08}, is a second-generation high-contrast imaging instrument for the Very Large Telescope (VLT), which finished its commissioning phase in 2014. Since then it has been routinely delivering unprecedented science results in the fields of dual-band imaging, differential polarimetry, and integral field spectroscopy of directly-imaged protoplanetary disks and young exoplanets (e.g. \citealp{Vigan:16a,Maire:16,Zurlo:16,Bonne:15,deBoer:16,Ginski:16}). The extreme adaptive optics (XAO) system of SPHERE, SAXO \citep{Sauvage_JATIS:16,Fusco:14,Fusco:16}, is capable of routinely achieving Strehl ratios of 90\% in the H-band. When this performance is combined with coronagraphic observation modes and optimised reduction pipelines, it is possible to achieve $5\sigma$ planet-star companion detectability ratios of better than $10^{-6}$ beyond angular separations of 375~mas \citep{Zurlo:16}.
	
\begin{figure}
	\hspace{-8pt}
	\includegraphics[width=0.5\textwidth]{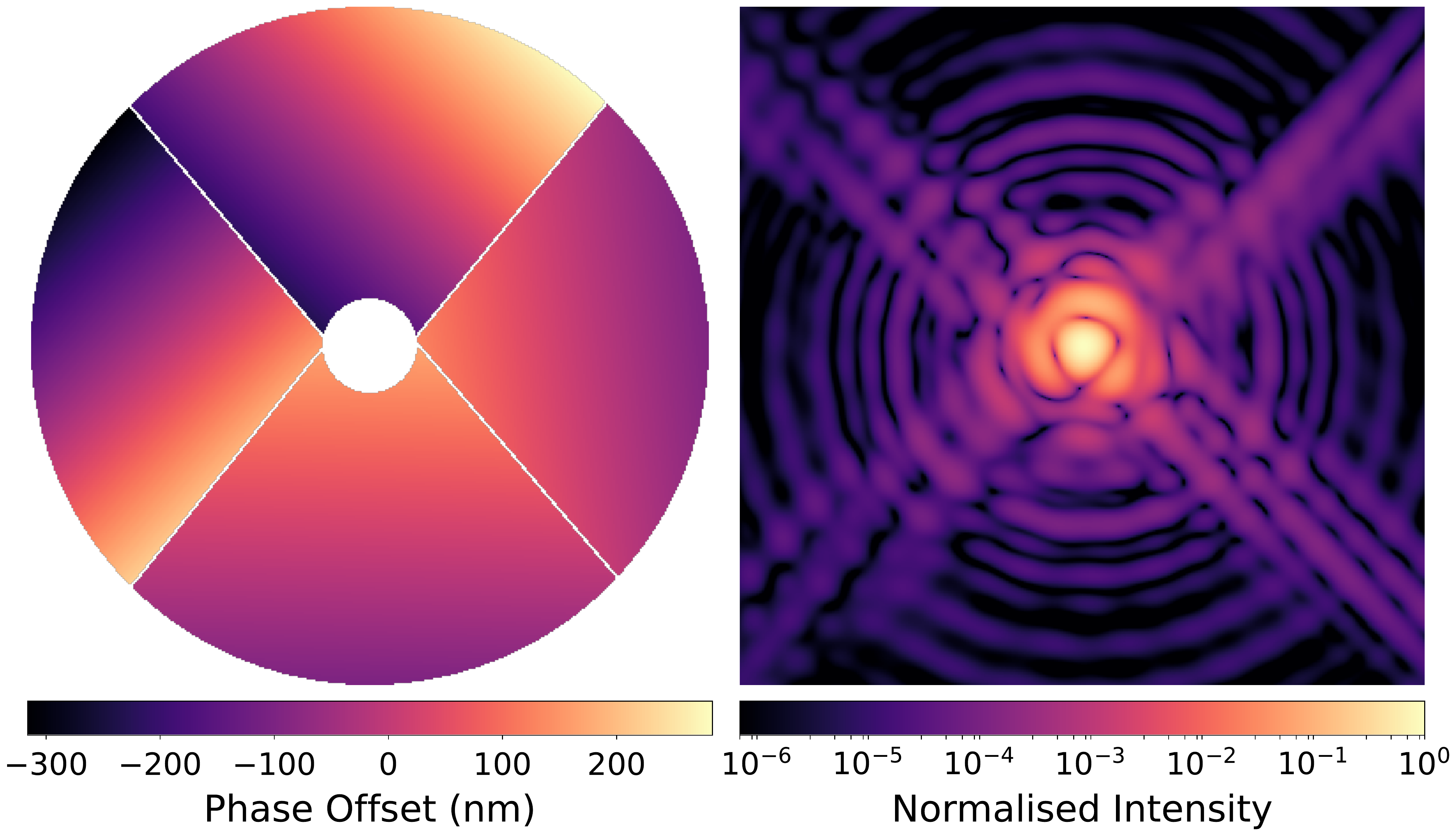}
	\caption{Typical example of the LWE phenomenon, based on on-sky measurements made with the ZELDA wavefront sensor. {\bf Left:} Parametrised PTT wavefront model based on a single ZELDA phase measurement. {\bf  Right:} The aberrated PSF corresponding to this LWE wavefront map, simulated at a wavelength of 1.536~${\rm\mu}$m. The PSF displays three notable side-lobes at the location of the first Airy ring (2.5~$\lD$), which correspond to the differential tip-tilt components seen across individual pupil segments in the aberrating phase map.}
	\label{Fig:1_LWEPSF}
	\end{figure}

	However, the instrument performance and ultimately the science yield of SPHERE is currently limited under the best observing conditions by the so-called low wind effect (LWE). This effect refers to a systematic degradation of the image quality of all three SPHERE detector arms (IRDIS, IFS, and ZIMPOL), which occurs when the wind speed at the altitude of the VLT dome drops below approximately $3~{\rm ms^{-1}}$ \citep{Sauvage:16}.	
	The characteristic LWE wavefronts consist of independent piston-tip-tilt (PTT) phase errors across one or more of the VLT pupil segments, and have been observed to reach up to 800~nm peak-to-valley error (PVE) on-sky as measured by a prototype of the Zernike wavefront sensor ZELDA \citep{NDiaye:14,NDiaye:16}. As shown in Fig.~\ref{Fig:1_LWEPSF}, this leads to a significant degradation of the imaging point-spread function (PSF) by creating multiple bright side-lobes at the location of the first Airy ring and increasing the amount of diffraction structure around the secondary mirror (M2) support spiders. 	
	This is an issue for both the coronagraphic and non-coronagraphic high-contrast observing modes of SPHERE, due firstly to increased photon noise and a lower Strehl ratio of off-axis companion sources. In reality the LWE is also a quasi-static phenomenon, and generates significant additional speckle noise on timescales and angular separations particularly detrimental to reference PSF subtraction and other high-contrast data reduction techniques, such as the angular differential imaging (ADI) and principal component analysis (PCA) classes of algorithm (e.g. \citealt{Marois:06,Lafren:07,Soummer:12,Amara:12}).
	
	This high-amplitude, quasi-static LWE may be considered a specific example of a more general `island effect' \citep{NDiaye:17}, which encompasses all differential PTT aberrations associated with pupil segmentation irrespective of underlying cause or temporal behaviour. Since examples of island effect behaviour are now also being reported intermittently at the SCExAO \citep{Jova:15} and GPI  instruments (\citealt{Mac:08}, V. Bailey, private communication, 2016), solutions developed for the LWE in SPHERE may well be applicable to similar issues faced by other instruments. This is expected to be especially important for the upcoming extremely large telescopes (ELTs), which will feature significantly more complex pupil geometries and may be correspondingly prone to these effects.
	
\begin{figure}
	\hspace{-8pt}
	\includegraphics[width=0.5\textwidth]{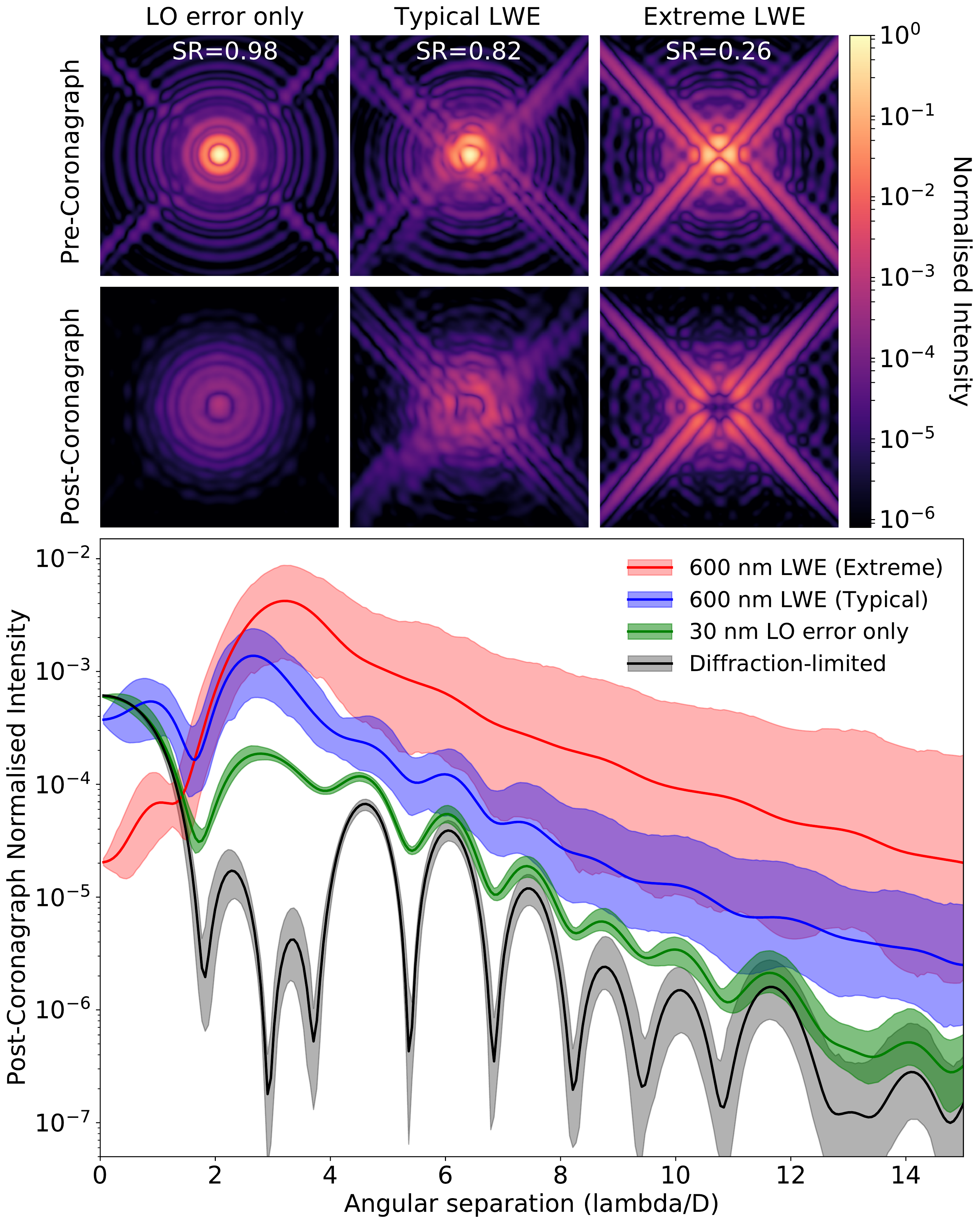}
	\caption{Simulated performance of the SPHERE APLC in the presence of various LWE models and low-order wavefront error, for an observing wavelength of $1.536~{\rm \mu m}$. {\bf Left image column:} LWE-free PSFs containing only 30~nm RMS of low-order aberrations with a $1/f^2$ spatial power spectrum, incoherently averaged over 100 random realisations. {\bf Centre column:} a typical three-lobed LWE model (600~nm PVE) identical to Fig.~\ref{Fig:1_LWEPSF}. {\bf Right column:} an extreme LWE with wavefront similar to a four-quadrant phase mask pattern (600~nm piston on two opposite VLT pupil quadrants). {\bf Top image row:} non-coronagraphic PSFs, including image Strehl ratio with respect to the diffraction-limited case. {\bf Bottom image row:} Corresponding on-axis coronagraphic PSFs. {\bf Main figure panel:} Radial average contrast curves of each post-APLC PSF, with shading denoting the 1$\sigma$ upper and lower bounds on azimuthal variation. Diffraction-limited performance in the absence of all aberration is shown by the black curve for comparison purposes.}
	\label{Fig:2_CoroEff}
	\end{figure}

	Figure~\ref{Fig:2_CoroEff} illustrates the degradation of raw contrast performance for the SPHERE apodised pupil lyot coronagraph (APLC) for two example LWE cases drawn from the SPHERE user manual \citep{ESO:v99}, using a Fourier propagation model of the three-plane coronagraph system described in \cite{Guerri:11}. It can be seen that diffraction-limited simulations (black curve, main panel) predict a raw contrast ratio of significantly better than $10^{-4}$ between 2-4~$\lD$, however this is not representative of real systems containing sources of non-common path error (NCPE). In order to provide a more realistic performance estimate, each PSF in Fig.~\ref{Fig:2_CoroEff} (and the corresponding green, blue, and red curves in the lower panel) includes the incoherent average of 100 random realisations of low-order, low-amplitude wavefront aberrations. These low-order wavefronts are created by drawing random Zernike mode coefficients, with the resulting phase maps then spatially filtered in the Fourier domain to have a $1/f^2$ decreasing spatial power spectrum often used to model NCPEs \citep{Sauvage:07,Lamb:16}. These are then scaled to have a 30~nm root-mean-square (RMS) error, representative of the calibration accuracy achieved in SPHERE after baseline NCPE calibration routines \citep{Fusco:14}. From this it can be seen that for a typical LWE amplitude of 600~nm PVE there is an increase in off-axis transmission of the central source of an order of magnitude with respect to the NCPE-limited case between 2-4~$\lD$. This alone would result in a factor of three increase in photon noise in the final reduced image at these angular separations, notwithstanding the inevitable impact of speckle variability due to a quasi-static LWE on the ultimate achievable contrast.
	
The current working hypothesis is that the LWE is caused by slow laminar airflow across the deep but narrow VLT M2 support spiders, which allows time for significant thermal exchange to occur \citep{Sauvage:15}. This results in sharp temperature changes and hence variations in the optical path depth of the air column across the width of each spider, thus generating discontinuities in phase such as that illustrated in Fig.~\ref{Fig:1_LWEPSF}a. This optical path depth hypothesis is supported by ESO simulations \citep{Sauvage:16}, which can reproduce the strength and overall morphology of the various wavefronts associated with the effect under realistic dome conditions. These characteristic wavefronts are however not seen in either SAXO Shack-Hartmann wavefront sensor (SH-WFS) data or in the deformable mirror (DM) actuator voltages during on-sky operation when the effect is present. This implies that the AO system is at best blind to this class of wavefront error, and at worst may be partially responsible for creating the effect due to unreliable sensing of phase discontinuities near the spiders. For this reason a complete understanding of the LWE, and other instances of the island effect, remains an active area of investigation with potentially significant implications for the design of future high-contrast imaging instruments.
	
	Attempts to eliminate the LWE phenomenon via dynamic control of telescope dome conditions (including increased ventilation, temperature control, and telescope pointing with respect to the prevailing wind direction) or manipulation of the AO closed-loop parameters have so far proved unsuccessful in reducing the strength or occurrence rates of the effect. Current efforts are ongoing to improve the thermal properties of the spiders by directly applying coatings with improved near-infrared (NIR) emissivity (M. Kasper, private communication, 2016), which if successful would allow the structure to better remain in thermal equilibrium with the surrounding air and thereby prevent phase discontinuities from arising.
	Another approach is to directly sense and compensate the LWE wavefront in real-time by introducing an additional wavefront correction loop into the SPHERE instrument: the ZELDA wavefront sensor has been shown to be an accurate truth sensor for the LWE during trials at the VLT in 2016, however until an upgrade to the SPHERE instrument can be performed ZELDA must convert the IRDIS focal plane into a pupil-plane sensor, preventing it from being used simultaneously with NIR science observations.
	
	This paper proposes an immediately implementable solution to directly sense and compensate the LWE, by using phase diversity techniques to turn the existing differential tip-tilt sensor (DTTS) camera, used for centring the stellar PSF on the NIR coronagraph \citep{Baudoz:10}, into a focal-plane wavefront sensor. This is an attractive solution as it requires no additional hardware or modification to the operation of existing SPHERE subsystems, as the correction commands may be applied as reference slope offsets to the main SAXO SH-WFS and therefore should be able to operate in parallel with the atmospheric XAO loop without conflict.
	For this task we propose the Fast \& Furious (F\&F) modified sequential phase diversity algorithm \citep{Keller:12,Kork:14}, so named because it uses a simplified model of the imaging system to obtain an analytical, computationally efficient phase reconstruction procedure (see Sect.~\ref{Sec:T_FF} for details).
	F\&F is also capable of using its own phase correction update cycle to provide the necessary phase diversity for complete focal-plane wavefront retrieval. This is in contrast to the majority of phase diversity approaches \citep{Gons:01,Sauvage:07,Lamb:16}, which require the periodic application of large controlled probe phases in order to reconstruct the aberrating wavefront. By eliminating this requirement, F\&F has the major advantage of being able to run continuously in closed-loop without degrading or interrupting science observations, enabling continuous real-time wavefront control.
	
	Fast \& Furious has been successfully tested in proof-of-concept simulations under imaging conditions emulating the SPHERE DTTS \citep{Wilby:16a} and in a general laboratory environment not specific to the LWE \citep{Kork:12,Kork:14}. While these preliminary simulations indicated that the algorithm should be robust against operating with the DTTS camera, before this solution may be implemented on-sky it is essential to verify that this performance is reflected in an appropriate laboratory environment. 
	In this paper we therefore present the results of LWE-specific lab testing using the Marseille Imaging Testbed for HIgh Contrast (MITHIC) \citep{Vigan:16b}, located at the Laboratoire d'Astrophysique de Marseille (LAM), the results of which are combined with improved closed-loop simulations to evaluate the potential performance of F\&F on-sky in SPHERE at the VLT.
	
	This paper is divided into the following sections: Sect.~\ref{Sec:Method} outlines the principle of the F\&F algorithm and presents details of both the MITHIC test-bench environment and supporting simulation tools. Sect.~\ref{Sec:Results} presents the main laboratory  closed-loop results and compares these to simulated performance predictions, and investigates the stability of F\&F at extremely low image signal-to-noise ratio (S/N). The limiting factors and lessons learned from this investigation are discussed in Sect.~\ref{Sec:Disc}, and final conclusions are drawn in Sect.~\ref{Sec:Conc}.

\section{Methodology}
\label{Sec:Method}

	\subsection{The Fast \& Furious algorithm}
	\label{Sec:T_FF}

	The F\&F algorithm refers to a sequential phase diversity technique based on \cite{Gons:02}, which has been extended to improve dynamic range and stability. It is capable of performing real-time wavefront phase retrieval when provided with a time-series of non-coronagraphic, narrowband focal-plane images and knowledge of the frame-to-frame phase commands applied by deformable elements in the system. Wavefront reconstruction is achieved by solving an analytical approximation to the stellar PSF in terms of the even and odd focal-plane intensity distributions, corresponding to the Fourier symmetries of the wavefront to be sensed. Phase diversity information is used only to break a sign ambiguity associated with calculating the even wavefront, which is most effectively provided by the phase correction command from the preceding iteration of F\&F.
	This has the major advantage that the sequential phase diversity process continually improves wavefront quality, allowing it to operate in parallel with continuous science observations. The algorithm is also highly computationally efficient, requiring only a single complex Fourier transform per iteration plus a small number of linear operations on image data. The correction cadence of F\&F will therefore be limited by the imaging camera readout frequency in most practical applications, whereas other phase diversity approaches are limited by the (significantly lower) frequency of phase probe injection.
	Unlike classical phase diversity however, F\&F is not capable of performing one-shot phase retrieval and must be operated in closed-loop for full wavefront compensation. Despite this it is always possible to reconstruct the odd wavefront component from any single PSF image, and with an appropriate choice of initial conditions it is almost always possible to achieve a systematic improvement in wavefront quality even on the first iteration.	
	
	A full derivation and analysis of the numerical properties of the F\&F algorithm can be found in \cite{Keller:12} and \cite{Kork:14}; a summary of the key equations necessary to implement the algorithm is presented here for reference. In the following description capitalised variables are used to denote two-dimensional pupil-plane quantities and lower-case variables denote two-dimensional focal-plane quantities, unless otherwise noted. The PSF $p = |\FT[Ae^{i\Phi}]|^2$ of an aberrated stellar image may be Taylor expanded to second order as a function of the even and odd focal-plane electric fields as
	\begin{equation}
	p \approx Sa^2 + 2a(ia\ast\phi_o) + (ia\ast\phi_o)^2 + (a\ast\phi_e)^2 ,
	\label{eq:PSF_approx}
	\end{equation}
	where $a=\FT[A]$ is the complex Fourier transform of the telescope aperture function $A$, which is assumed to be real and symmetric, while $\phi_o = \FT[\Phi_o]$ and  $\phi_e = \FT[\Phi_e]$ are the complex Fourier transforms of the odd and even components of the wavefront phase map $\Phi$, with $\Phi_o=-\Phi_o^{\rm T}$ and $\Phi_e = \Phi_e^{\rm T}$, such that $\Phi = \Phi_o + \Phi_e$. The scalar normalisation factor $S = (1-\sigma_\phi^2)$ is approximately equal to the Strehl ratio of the most recent image, where $\sigma_\phi^2$ is the total wavefront variance: this is effectively the first-order Taylor expansion of the Mar{\'e}chal approximation \citep{Lewis:04}.  Here $\FT$ is the Fourier transform operator, and $\ast$ is the convolution operator. 
	This formulation takes advantage of the enforced symmetry properties of all pupil-plane quantities to simplify the expressions, and to remain consistent with the symmetry properties of the complex Fourier transform.

	The expression above may be more conveniently expressed by defining the odd and even focal-plane fields as the two real quantities
	\begin{align}	
	\label{eq:y}
	y &= i\FT[A\Phi_o] = (ia\ast\phi_o),~{\rm and} \\
	v &= \FT[A\Phi_e] = (a\ast\phi_e),
	\label{eq:v}
	\end{align}
	leading to analytical solutions for the complete odd field and the absolute value of the even field, by separating Eq.~\ref{eq:PSF_approx} according to symmetry and solving. This yields
	\begin{align}	
	\label{eq:yv_sol}
	y &= ap_o/(2a^2 + \epsilon),~{\rm and}\\
	|v| &= \sqrt{|p_e - (Sa^2 + y^2)|} ,
	\label{eq:v_sol}
	\end{align}
	where $p_o$ and $p_e$ are the odd and even components of $p$  by direct analogy with $\Phi_o$ and $\Phi_e$. Here the scalar $\epsilon$ parameter is introduced as a method of regularisation for pixels where $a$ tends towards zero, which is typically set to a factor of ten above the noise threshold in the image sequence.
	
	In order to estimate the signs of the even focal-plane field, it is necessary to introduce a second PSF image which differs from the first by a known phase offset $\Phi_d$, such that
	\begin{alignat}{4}
	p_1 &= Sa^2 + ~~~~~2ay && + ~~~~~y^2 && + ~~~~~v^2,~{\rm and} \\
	p_2 &= Sa^2 + 2a(y+y_d) && + (y+y_d)^2 && + (v+v_d)^2,
	\label{eq:PSF_div}
	\end{alignat}
	where $v_d$ and $y_d$ correspond to this additional phase diversity between frames by analogy with Eqs.~\ref{eq:y}~\&~\ref{eq:v}. Solving for $v$ yields an independent expression in terms of the even components of each PSF, $p_{1e}$ and $p_{2e}$, however this solution is extremely prone to noise due to the subtraction of two similar PSFs. It is therefore significantly more robust to take only the signs of this estimate,
	\begin{equation}
	{\rm sign}(v) = {\rm sign}\left(\frac{p_{2e} - p_{1e} - (v_d^2 + y_d^2 + 2yy_d)}{2v_d}\right),
	\label{eq:v_signs}
	\end{equation}
	and combine these with the magnitude $|v|$ computed from Eq.~\ref{eq:v_sol}.
	
	The final estimate of the wavefront phase may then be reconstructed via a single complex inverse Fourier transform, by taking into account the symmetries of the odd and even focal-plane fields to give
	\begin{equation}
	A\Phi = \FT^{-1}[{\rm sign}(v)|v|-iy] .
	\label{eq:Finaleq}
	\end{equation}
	This estimate may then be applied using a wavefront correcting element, which in turn becomes the new phase diversity $\Phi_d$ for the next iteration. Since ${\rm sign}(v)$ is undefined on the first iteration where phase diversity information is not available, it is optimal in the case of small wavefront aberrations to choose ${\rm sign}(v_0) = {\rm sign}(\FT[A]) $. 
	
	Key assumptions made implicitly by F\&F include the use of a symmetric telescope pupil function, the presence of phase-only aberrations and a monochromatic light source. An improved version of the algorithm dubbed FF-GS, which includes Gerchberg-Saxton style iterative steps, has been developed \citep{Kork:14} which can overcome these first two limitations by also enabling amplitude retrieval. It was not however found to be necessary to implement FF-GS in this work, due to the near-symmetric SPHERE pupil generating only small systematic reconstruction errors which may be removed by spatial filtering, therefore not warranting the increased complexity and lower stability of FF-GS compared to F\&F alone.

	\subsection{Implementing F\&F in SPHERE-like simulations}
	\label{Sec:M_Sim}
	
	The first step towards using F\&F as a LWE-compensator for SPHERE is to simulate as closely as possible the observing conditions of the DTTS camera, and verify that the algorithm is both efficient at eliminating LWE wavefronts and stable during continuous operation in the absence of the effect. The DTTS Hawaii I camera is capable of operating at high frame-rates (1~Hz - 1~kHz) with only a $32\times32$~pixel field of view at 3.5 pixels per $\lD$ sampling, with images stacked to provide tip-tilt correction at a cadence of 1~Hz. This is operated at a wavelength of $1.536~{\rm\mu m}$ (H-band) with a 3\% bandwidth and is situated behind a 2\% beam-splitter to avoid unnecessary science throughput losses, hence DTTS images normally have low S/N. F\&F must therefore first and foremost be robust against dominant detector noises sources and a limited image size. Other considerations include the presence of amplitude variations, most notably the presence of the SPHERE NIR coronagraph amplitude apodiser (located upstream of the DTTS beam-splitter) and pupil rotation, but also errors in representing wavefront control commands due to fitting errors associated with a finite DM resolution, as well as systematic errors in applying reference offsets on an existing AO loop.
	
	A dedicated python simulation package has been developed for the purpose of validating F\&F under SPHERE-like conditions, which is also used throughout this work. This code is capable of generating realistic DTTS-like image sequences using an XAO-corrected turbulence phase-screen simulator and quasi-static NCPE model with appropriate coherence timescale and spatial power spectrum, with photon and detector noise sources added to achieve the desired S/N. 
	A comprehensive overview of the main code features is provided in \cite{Wilby:16a}, which also reported initial simulation results demonstrating that F\&F should be capable of providing significant and robust improvements in wavefront quality for the case of the LWE. The key steps used in this implementation F\&F are described below, which also apply to the code used for MITHIC laboratory testing (see Sect.~\ref{Sec:M_lab}).
	
	As part of the preliminary data reduction step, a windowing function was applied to images in order to suppress pixel-to-pixel noise by removing the high spatial frequencies containing no retrievable wavefront information. This took the form of a radial sigmoid low-pass spatial filter, with a radial cut-off determined by image S/N: this step is described further in Sect.~\ref{Sec:S/N}.
	Two consecutive images $p_1$ and $p_2$ were then used along with the appropriate phase diversity command $\Phi_d$ to calculate the F\&F phase reconstruction $A\Phi$, as described by Eqs.~\ref{eq:yv_sol},~\ref{eq:v_sol},~\ref{eq:v_signs}~\&~\ref{eq:Finaleq}. 
	The maximum spatial frequency which can be controlled by F\&F is ultimately set by the field-of-view (FOV) of the input images, and the Fourier transform operation must be constructed so as to ensure that the reference field $a$ has both the same focal-plane sampling as the input images and at least the same FOV. This may be achieved by using a discrete fast Fourier transform (FFT) operation with an appropriately zero-padded aperture function $A$, but to fully optimise the algorithm a non-FFT based method such as the semi-analytical approach described in \cite{Soummer:07} should be used. This allows for an arbitrary focal-plane sampling and FOV whilst computing $A\Phi$ at an appropriate resolution for the deformable element being used, thereby maximising computational speed.
	
	In order to further suppress pixel-to-pixel noise and systematic artefacts associated with asymmetric pupil features (such as spiders), the raw F\&F wavefront estimates were projected onto a low-order modal correction basis, which can also be customised to constrain the degrees of freedom that F\&F is able to control. The most appropriate modal correction basis to use given the current model of LWE wavefront morphology is a segmented basis consisting of independent PTT components for each VLT pupil quadrant, although Zernike or disk harmonic modes are less sensitive to specific pupil orientation and would also allow F\&F to correct additional errors including NCPEs. 
	Finally, the resulting phase maps were spatially filtered with a Gaussian kernel to approximate the limited spatial frequency response of the $41\times41$ actuator SPHERE DM. For simplicity it was assumed that these final filtered wavefront estimates could be applied accurately on the DM via reference offsets to the zero points of the main SAXO AO loop. Wavefront corrections were therefore directly applied on the (simulated) deformable element throughout this work. This assumption is encouraged by successful preliminary testing of the SPHERE DM response to piston offset commands \citep{Sauvage:16}, but an end-to-end simulation involving the SPHERE filtered SH-WFS and DM interaction matrix is beyond the scope of this paper.
	
	More detailed and diagrammatic explanations of F\&F wavefront correction loop architectures can be found in \cite{Kork:14} and \cite{Wilby:16a}. The code implementation described in the latter paper has since been enhanced to explore additional considerations for SPHERE performance, with the capability to apply both systematic and random temporal variations in the gain of individual DM actuators (and hence imperfect reproduction of F\&F phase estimates), the capacity to add additional amplitude aberrations and pupil rotation, a more realistic spatial filtering window function for atmospheric phase-screen generation, and a $1/f^2$ filtered spatial power spectrum of injected NCPEs. This allows an extension of the already published simulation results, especially for addressing the impact of factors such as amplitude aberrations or reconstruction errors when applying F\&F update commands through a real AO system.
	
	\subsection{Laboratory verification of F\&F on the MITHIC bench}
	\label{Sec:M_lab}
	
	In order to verify whether the simulation package discussed above is accurate in its implementation of F\&F and in its treatment of the most important factors for on-sky observation, we performed closed-loop tests with the MITHIC high-contrast testbench at LAM \citep{Vigan:16b}. The F\&F code package was implemented on the bench as described in the previous sub-section and executed for a number of controlled convergence tests, with variable parameters including the amplitude and type of wavefront error, image FOV, pupil apodisation, and image S/N. Each set of laboratory conditions was also run through an equivalent closed-loop simulation to provide a direct means of comparison between the two approaches.	The various parameters and results of these tests are listed in Table~\ref{Tab:1_LabResults}, and are discussed in detail in Sect.~\ref{Sec:LabvSim}.
	
	\begin{figure}
	\hspace{-12pt}
	\centering
	\includegraphics[width=0.51\textwidth]{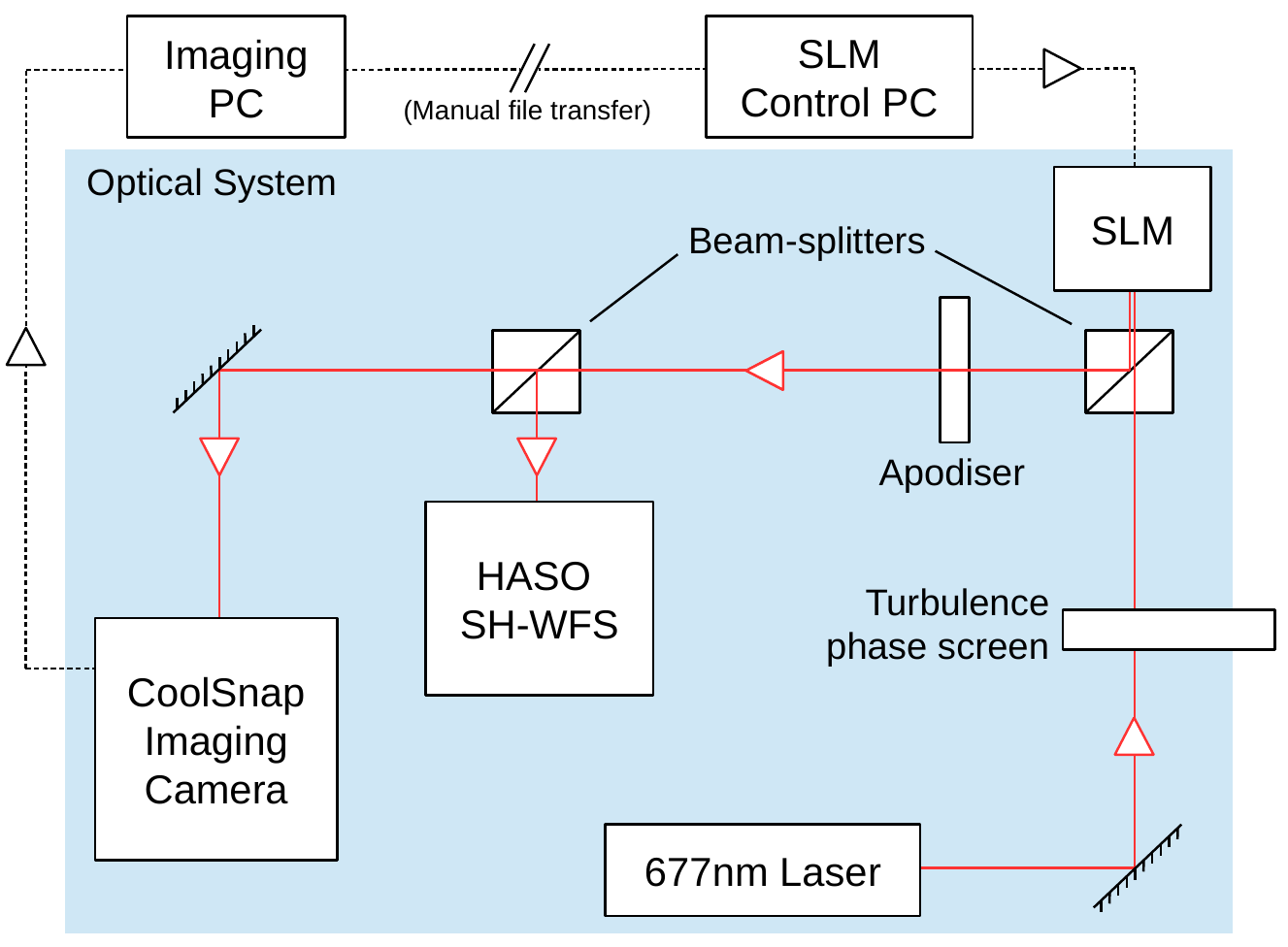}
	\caption{ Schematic of the MITHIC bench configuration used for this investigation. The spatial light modulator (SLM) is placed in a face-on reflective configuration via a double pass through a beam-splitter. 
	Images taken via the CoolSnap camera GUI were manually transferred from the imaging PC to the SLM control PC via a separate LAM server, before performing F\&F wavefront reconstruction as described in Sect.~\ref{Sec:M_Sim} and applying the new wavefront commands on the SLM via a custom GUI. Each closed-loop iteration of F\&F typically took between 30~s and a minute to complete, of which the F\&F code runtime was a negligible fraction. 
	The HASO SH-WFS was used only for the pre-compensation of MITHIC bench alignment errors and was not operated during F\&F closed-loop tests. The turbulence phase screen was aligned either on the static LWE pattern or a clear aperture, and was not used to simulate dynamical turbulence. }
	\label{Fig2b:MITHIC}
	\end{figure}
	
	The MITHIC bench operates at visible wavelengths (677~nm) and is optimised for high-contrast coronagraphic imaging using the Roddier and Roddier phase mask coronagraph \citep{Roddier:97,NDiaye:11}, with the capability for pupil-plane wavefront measurement provided by Shack-Hartmann and ZELDA wavefront sensors, in addition to the COFFEE coronagraphic phase diversity estimator \citep{Paul:14} for focal-plane NCPE control. Figure~\ref{Fig2b:MITHIC} illustrates the optical layout of the bench as was used for this investigation, which is similar to that described in \cite{Paul:14}. A newly-installed phase-screen turbulence simulator \citep{Vigan:16b} can be used to inject either single layer dynamic turbulence or a variety of static wavefront error patterns into the beam, including a LWE aberration with the same morphology as in Fig.~\ref{Fig:1_LWEPSF} and an estimated amplitude of 98~nm PVE. Wavefront control was achieved using a Hamamatsu liquid-crystal-on-silicon spatial light modulator (LCOS-SLM), which is situated behind a beam-splitter in a face-on reflective configuration and samples the re-imaged pupil with 273 pixels across the diameter. The CoolSnap HQ2 1392~$\times$~1040 pixel interline CCD camera was used for final focal-plane imaging, sampling the final PSF at 9.6~pixels per $\lD$. For this work the camera was used in non-coronagraphic imaging mode, with the images numerically binned to a resolution of 3.3~pixels per $\lD$ and cropped to a $32\times32$ pixel FOV, matching the default imaging parameters of the DTTS. The PSF S/N was adjusted by varying the laser input power for a fixed exposure time of 1~ms, and was calculated based on the central image pixel with respect to the background noise floor.

	In the absence of artificially injected sources of wavefront error, the MITHIC bench was measured by the HASO-3 SH-WFS to contain 96~nm RMS of astigmatism-dominated low-order static error, due to optical mis-alignments. After wavefront flattening using the SLM it was estimated that residual aberrations were reduced to the level of 10~nm RMS, with the imaging camera PSF displaying four consecutive unbroken Airy rings. This phase correction was then manually applied as a flat wavefront command using the SLM during this investigation, ensuring that injected aberrations comprised the vast majority of total wavefront error in the system for closed-loop tests.
	
	\begin{figure}
	\hspace{-10pt}
	\includegraphics[width=0.51\textwidth]{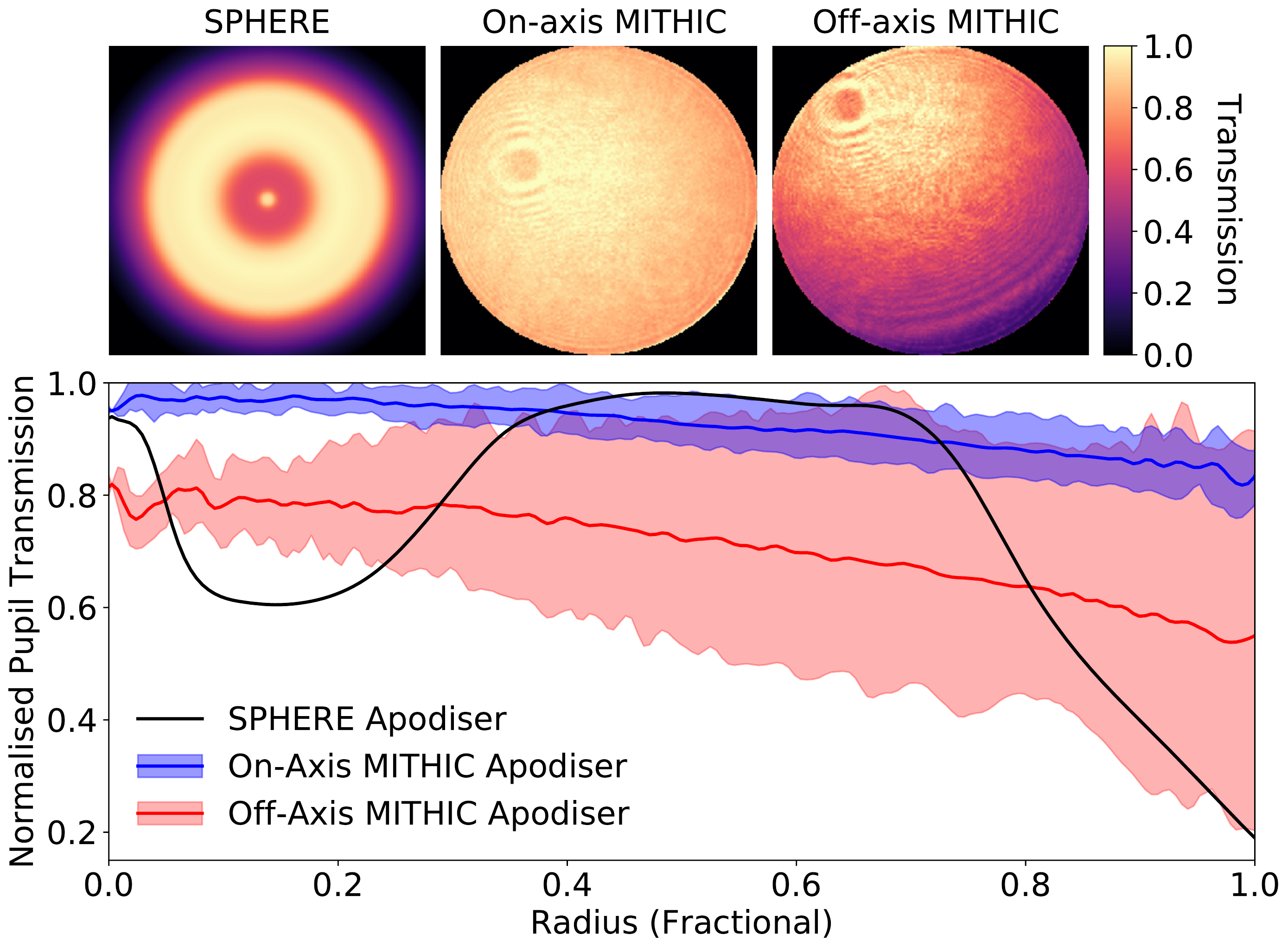}
	\caption{Throughput images (top panels) and radial transmission profiles (lower panel) of the amplitude apodisers relevant to this investigation. {\bf Top row, from left:} SPHERE APLC apodiser model of \cite{Guerri:11}, and pupil-imaging measurements of the on-axis and off-axis transmission (respectively) of the weak MITHIC bench apodiser. {\bf Lower panel:} Dark lines denote the mean radial throughput curve of each transmission profile, with shading illustrating the maximum and minimum bounds in the azimuthal direction.}
	\label{Fig:3_Apods}
	\end{figure}

	A weak amplitude apodiser was included in the setup for a small subset of tests, inserted in the pupil shortly after the SLM in both an on-axis and off-axis position as shown in Fig.~\ref{Fig:3_Apods} in order to test the response of F\&F to amplitude effects. During most SPHERE coronagraphic observations the DTTS imaging path also includes the strong APLC apodiser shown in black in Fig.~\ref{Fig:3_Apods}, but a similar apodiser was not available in MITHIC for these tests. This is not in itself a concern for F\&F performance: the radial amplitude function of the APLC apodiser is completely symmetric, allowing it to be included when calculating the reference focal plane field $a$ which defines the zero point for F\&F wavefront reconstruction, without violating any of the assumptions on pupil geometry made in Sect.~\ref{Sec:T_FF}. The important consideration which is investigated here is whether the stability of F\&F is limited by unknown amplitude errors, for example due to optical mis-alignments or errors in the pupil model used for F\&F. The MITHIC apodiser pupil functions were therefore not provided to F\&F when generating the reference field $a$ for these tests, which instead used a uniform circular aperture matched to the MITHIC beam.
	
	Due to the practicalities of using separate GUI interfaces for imaging and SLM control, coupled with the non-networked computing architecture currently implemented in MITHIC, wavefront correction could only be achieved by manually closing the loop. This limited the number of iterations which could reasonably be performed to a maximum of 25, due to time restrictions and risk of human error. As simulations and previous laboratory tests of F\&F indicated that stable convergence is typically achieved in 5-6 iterations, this approach was deemed sufficient to characterise the initial convergence behaviour and place limits on the short-term post-convergence stability. 
	
	The limited number of closed-loop iterations, combined with the slow and irregular wavefront update frequency of manual control, also made it impractical to properly apply  wavefront dynamics representative of SPHERE on-sky observations. All injected LWE and low-order aberrations in this investigation were therefore static, whether applied on the SLM or via static patterns on the turbulence phase screen. Most importantly, this means that MITHIC images did not include a simulation of atmospheric turbulence, quasi-static NCPEs or variability of the LWE itself.
	However, none of these dynamical factors were found to limit F\&F performance when studied in the prior closed-loop simulations of \cite{Wilby:16a}, provided that the wavefront correction loop can run sufficiently fast compared to the LWE coherence timescale: this point is discussed further in Sect.~\ref{Sec:S/N}. The lack of an incoherent atmospheric speckle background does mean that the Strehl ratios quoted in this paper are close to the diffraction limit, and must be scaled down by the typical H-band performance of SAXO in order to make a comparison with expected on-sky performance.
	
	In future it would be advantageous to fully automate the MITHIC wavefront correction loop, which would eliminate user error and allow more thorough investigation into the long-term closed-loop stability of focal-plane sensors, and their correction cadence requirements under variable conditions more closely resembling on-sky observations.

\section{Results}
\label{Sec:Results}

	\subsection{Comparison of MITHIC bench and simulation results}
	\label{Sec:LabvSim}	
	
	\begin{figure*}
		\hspace{-8pt}
		\includegraphics[width=1.0\textwidth]{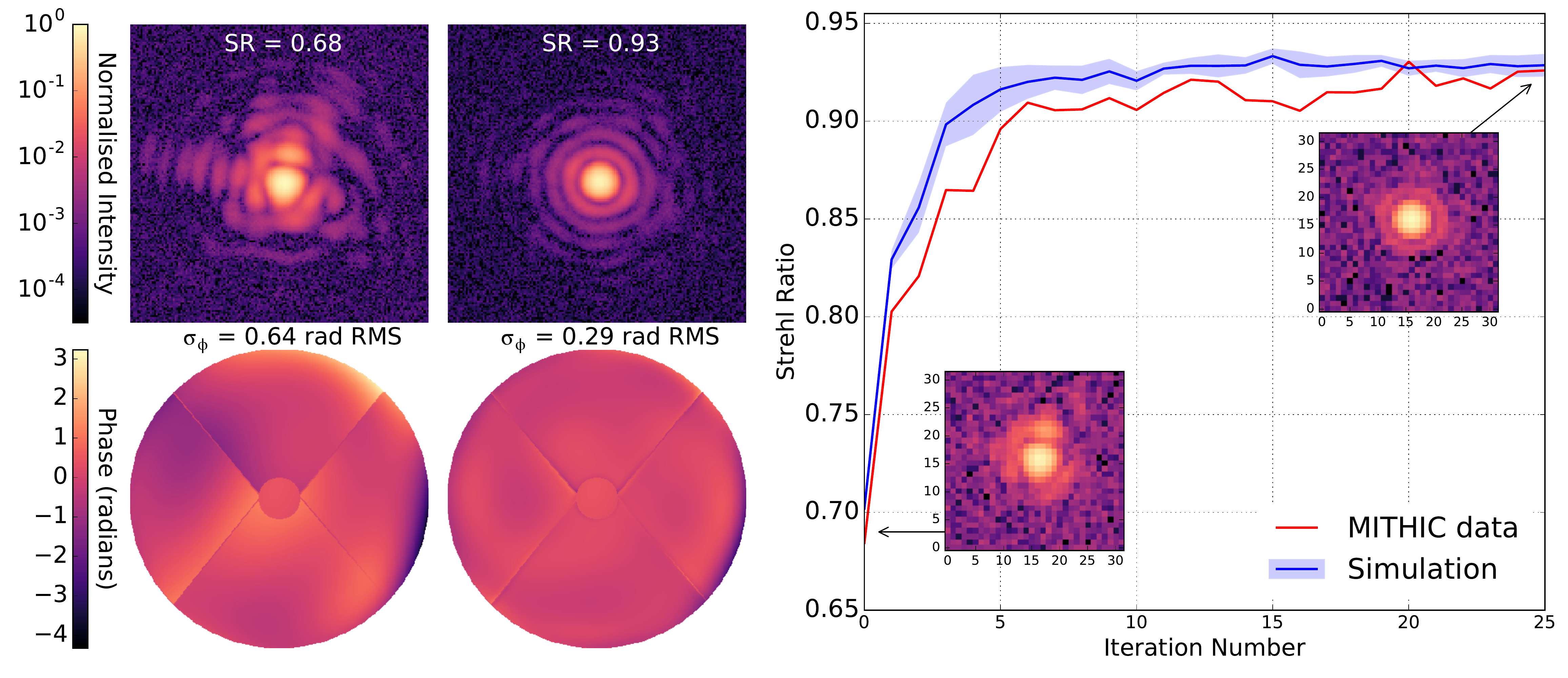}
		\caption{Comparison of F\&F convergence behaviour between MITHIC laboratory data and equivalent simulations (see Table~\ref{Tab:1_LabResults}, row 1 for details). {\bf Left:} High-resolution, high-S/N MITHIC focal-plane images (top) and corresponding residual wavefront error maps (bottom), before and after F\&F correction (first and second columns respectively). High-frequency residuals visible in the final wavefront error map are dominated by fitting error enforced by the DM-like filtering of wavefront corrections. {\bf Right:} Plot of estimated image Strehl ratio as a function of closed-loop iteration number, showing close agreement between a single convergence of MITHIC data (red) and an average of ten simulations (blue), with shading denoting the 1$\sigma$ limit. The two inset images show the $N=0$ and $N=25$ DTTS-like input focal-plane images provided for F\&F for wavefront reconstruction.}
		\label{Fig:4_LabVer}
		\end{figure*}

		Figure~\ref{Fig:4_LabVer} shows the results of the most important convergence test performed on MITHIC, the details of which are listed in the first row of Table~\ref{Tab:1_LabResults}. This test applied a static 319~nm PVE LWE wavefront phase pattern on the SLM, which was chosen to provide half a wave of error at the 677~nm MITHIC laser wavelength. For the purposes of F\&F this is directly equivalent to solving a 724~nm PVE LWE in the H-band, which is close to the strongest LWE amplitudes regularly seen on-sky with SPHERE. The setup did not contain a SPHERE aperture mask but used a clear circular pupil: regions of LWE wavefront phase corresponding to the VLT M2 obscuration and support spiders were therefore set to zero phase.
		In addition to the LWE, 39~nm RMS of low-order aberrations were also applied on the SLM, with coefficients randomly drawn from the first 19 Zernike modes. The image S/N and 32~x~32 pixel FOV were chosen to be representative of DTTS images, and a weak on-axis apodiser was included in the beam. Corrections were made on the SLM by first projecting the output of F\&F onto a combination of the segmented PTT mode basis and a 50-mode low-order Zernike basis, and then spatially filtering to mimic the finite actuator influence function of the $41~\times~41$ actuator SAXO DM. This was achieved by convolving the resulting phase map with a Gaussian of full-width half-maximum equal to two thirds of the SAXO actuator spacing, which is qualitatively representative of the resolution seen in reference slope control tests by \cite{Sauvage:16}. By not applying phase corrections at the full $273~\times273$ pixel resolution of the Hamamatsu SLM with which the aberrating phase was initially implemented, we mimicked the interplay between a finite resolution DM and non-discrete upstream aberrating wavefront, thereby providing a better estimate of the correcting power of the SPHERE DM.
		
		The top-left image row shows initial and final PSFs taken at high S/N, showing the clear improvement in Strehl ratio from 68\% to 93\% associated with the elimination of LWE and the majority of the low-order wavefront within five closed-loop iterations. The residual wavefront error maps show that the uncorrected wavefront error was dominated by high-frequency components along the edges of the spiders and along one edge of the pupil. This was a result of the DM-like spatial filtering of the correction wavefront, and is representative of the capabilities of such an implementation on-sky.
		
		In the right-hand panel, the image Strehl ratio is directly compared for each iteration between MITHIC data and an average of ten simulations matching the laboratory conditions. Multiple simulations were run in order to place a limit on the reproducibility of convergence under noisy conditions, with only the individual pixel-to-pixel noise allowed to vary between realisations. This plot shows that there is an extremely close match between predicted and obtained performance in this scenario, with very similar final Strehl ratios achieved and laboratory convergence only very slightly lagging behind the simulated curve. It was also estimated from SH-WFS measurements during this convergence process that there existed approximately 10~nm RMS of residual NCPE in the system, for which F\&F is in principle also capable of correcting. This means that the laboratory implementation was dealing with additional wavefront error when compared to the simulated case, which may explain this small decrease in convergence efficiency. 
		
		 All Strehl ratios for MITHIC bench images quoted in this paper were estimated according to a modified encircled energy metric described by
		\begin{align}
		\label{eq:SR}
		S_{\rm MITHIC} \approx&~{\rm EE}(p_{\rm MITHIC})\frac{S_{\rm sim}}{{\rm EE}(p_{\rm sim})}~, {\rm where} \\[10pt]
		{\rm EE}(p) =&~\frac{\sum p(r<1.22\lambda/D)}{\sum p(r<3.5\lambda/D)}
		\label{eq:EE}
		\end{align}
		is the ratio of encircled energies between the Airy core and first Airy ring of the given PSF $p$, which provides an estimate of image quality using the regions visible above the noise floor. As the encircled energy metric systematically over-estimates the true Strehl ratio of MITHIC images $p_{\rm MITHIC}$ by ignoring high-frequency aberrations, the second term in Eq.~\ref{eq:SR} used to account for this bias using the known true Strehl ratio $S_{\rm sim}$ and encircled energy of the simulated PSF $p_{\rm sim}$ of matched simulations. Such a correction was possible due to the close consistency between laboratory and simulation results, such that the correction factors could be expected to be very similar for the two cases. This metric was chosen over more readily available metrics such as using the SLM residual wavefront map or a reference image PSF from the MITHIC bench, as these can be biased by factors such as residual NCPEs or intensity variations in the system, and in practice were seen to occasionally predict Strehl ratios of greater than one.
		
		\begin{table*}
\begin{adjustwidth}{0cm}{}
\caption{Summary of the key parameters of all laboratory F\&F convergence tests performed on the MITHIC bench. The first row corresponds to the most challenging scenario presented in Fig.~\ref{Fig:4_LabVer}. Strehl ratio (SR) estimates are calculated normalised encircled energy measurements of the PSF core and first Airy ring, as described by Eq.~\ref{eq:SR}. Final RMS error is estimated from the residual wavefront map implemented on the SLM, or the Mar{\'e}chal approximation when significant non-SLM induced aberrations are involved. The quoted $1\sigma$ errors on S/N, Strehl ratio, and residual RMS are calculated from the final five frames in each test sequence, and as such are a reflection of post-convergence stability. 
The final three rows denote test cases attempting to force F\&F to avoid correcting specific wavefront errors, by manipulating the correction basis or subtracting reference offsets from F\&F output wavefronts.
}
\vspace{-4pt}
{\renewcommand{\arraystretch}{1.1}
\tiny
 \begin{tabular}{cccccccccc} 
 \hline
 \rowcolor{lightgray}{\bf Row} & {\bf LWE~(PVE / RMS)} & {\bf Low-Order~(RMS)} & {\bf Corr. Basis} & {\bf Apodiser} & {\bf FOV (px)} & {\bf Final S/N} & {\bf Initial SR} & {\bf Final SR} & {\bf Final RMS}\\ 
 \Xhline{1\arrayrulewidth}
 1. & 319 / 59~nm  & 39~nm (Zernike\textcolor{red}{$\bf^1$}) & PTT + Zern & On-axis & 32 & $445\pm2$ & 68~\% & $93\pm1$~\% & $32.2\pm0.5$~nm \\ 
 \rowcolor{vLightGray}
 2. & 319 / 59~nm & 0~nm & PTT + Zern & None & 32 & $382\pm1$ & 72~\% & $94\pm1$~\% & $20.5\pm0.8$~nm \\ 
 3. & 98 / 16~nm\textcolor{red}{$\bf^2$} & 96~nm (Bench\textcolor{red}{\bf$^3$}) & PTT + Zern & None & 32 & $6247\pm6$  & 56~\% & $91.5\pm0.5$~\% & $28\pm1$~nm\textcolor{red}{$\bf^4$} \\  
 \rowcolor{vLightGray}
 4. & 319 / 59~nm & 0~nm & PTT + Zern & Off-axis & 32 & $4400\pm30$ & 72~\% & $94\pm1$~\% & $24\pm2$~nm \\ 
 5. & 319 / 59~nm & 0~nm & PTT + Zern & On-axis & 32 & $6410\pm20$ & 74~\% & $97.1\pm0.5$~\% & $21\pm2$~nm \\
 \rowcolor{vLightGray}
 6. & 98 / 16~nm\textcolor{red}{$\bf^2$} & 0~nm & PTT + Zern & None & 32 & $6450\pm30$ & 94~\% & $96.2\pm0.1$~\% & $20.7\pm0.3$~nm\textcolor{red}{$\bf^4$}\\ 
 7. & 319 / 59~nm & 0~nm & PTT + Zern & None & 32 & $6380\pm40$ & 73~\% & $94.9\pm0.1$~\% & $18\pm1$~nm\\
 \rowcolor{vLightGray}
 8. & 319 / 59~nm & 0~nm & Zern & None & 32 & $5980\pm20$  & 73~\% & $91.1\pm0.1$~\% & $24.8\pm0.5$~nm\\ 
 9. & 319 / 59~nm & 0~nm & PTT~ & None & 32 & $5970\pm40$ & 73~\% & $94.5\pm0.2$~\% & $14.4\pm0.6$~nm\\ 
 \rowcolor{vLightGray}
 10. & 319 / 59~nm & 0~nm & PTT~ & None & 100 & $6070\pm30$  & 73~\% & $97.9\pm0.1$~\% & $14.9\pm0.5$~nm\\ 
 11. & 319 / 59~nm & 0~nm & PTT\textcolor{red}{$\bf^5$} & None & 100 & $6020\pm30$ & 73~\% & $98.0\pm0.1$~\% & $16.4\pm0.7$~nm\\
 \rowcolor{vLightGray}
 12. & 98 / 16~nm\textcolor{red}{$\bf^2$} & 0~nm & None & None & 100 & $5850\pm20$  & 93~\% & $97.1\pm0.2$~\% & $18.5\pm0.6$~nm\textcolor{red}{$\bf^4$}\\ 
 13. & 319 / 59~nm & 0~nm & None & None & 100 & $5820\pm30$ & 72~\% & $94.8\pm0.2$~\% & $22.3\pm0.7$~nm\\
 \Xhline{5\arrayrulewidth}
 \rowcolor{vLightGray}
 14. & 319 / 59~nm & 0~nm & PTT\textcolor{red}{$\bf^6$} & None & 32 & $5570\pm70$ & 74~\% & $86\pm1$~\% & $32\pm2$~nm\textcolor{red}{$\bf^7$} \\
 15. & 319 / 59~nm & 43~nm (Focus) & PTT\textcolor{red}{$\bf^8$} & None & 32 & $5430\pm40$ & 66~\% & $82\pm1$~\% & $42\pm1$~nm\textcolor{red}{$\bf^7$} \\ 
 \rowcolor{vLightGray}
 16. & 98 / 16~nm\textcolor{red}{$\bf^2$} & 77~nm (Zernike\textcolor{red}{$\bf^1$}) & PTT~ & None & 32 & $3790\pm20$  & 66~\% & $71\pm1$~\% & $64\pm1$~nm\textcolor{red}{$\bf^{4}$} \\
 \end{tabular}}\\ [5pt]
 
 \raggedright\tiny 
 \textcolor{red}{$\bf^1$} Low-order static wavefront, consisting of the first 19 non-trivial Zernike modes (excluding PTT) with randomly drawn coefficients. \newline
 \textcolor{red}{$\bf^2$} LWE aberration is applied via the weak static phase pattern included in the MITHIC turbulence module \citep{Vigan:16b}, instead of via the SLM. \newline
 \textcolor{red}{$\bf^3$} Residual astigmatism-dominated MITHIC bench alignment errors, which were measured by the HASO SH-WFS and pre-compensated for all other tests. \newline
 \textcolor{red}{$\bf^4$} RMS residuals are estimated via Mar{\'e}chal approximation, as the majority of wavefront error is not applied via the SLM and thus residual phase is poorly known. \newline
 \textcolor{red}{$\bf^5$} No DM-like spatial filter is applied to F\&F outputs, hence wavefront corrections are applied at the full SLM resolution.\newline
 \textcolor{red}{$\bf^6$} Global tip-tilt is subtracted from F\&F estimates before projection onto the PTT basis. The PSF is initially shifted by $1\times1$ pixels from the F\&F reference centroid. \newline
 \textcolor{red}{$\bf^7$} F\&F exhibits divergent behaviour within ten closed-loop iterations. \newline
 \textcolor{red}{$\bf^8$} 43~nm of focus is subtracted from each F\&F output wavefront estimate before projection onto the PTT basis, in an attempt to apply a zero-point offset.
 \label{Tab:1_LabResults}
 \end{adjustwidth}
\end{table*}

Table~\ref{Tab:1_LabResults} also summarises the laboratory performance obtained with F\&F in MITHIC for a range of additional key tests, which illustrate the behaviour of the algorithm under variable conditions. It can be seen that for all but the final three test cases F\&F returned the image Strehl ratio to over 90\%, a gain of typically greater than 20\% on the starting value. 
	The right-most column quotes the post-convergence RMS wavefront error, which was in most cases estimated directly from the last phase command applied on the SLM after closed-loop convergence. These RMS residuals were found to be typically on the same order as the 10~nm RMS MITHIC bench alignment residual after HASO pre-compensation, and agree well with the quoted Strehl ratios through the use of the Mar{\'e}chal approximation. In cases where the target wavefront error was not injected with the SLM, such as when using the turbulence screen LWE aberration or removing the MITHIC bench wavefront flat command, this approach is inherently biased. For these tests the estimate of the final RMS residual was then simply made using the Mar{\'e}chal approximation with the final image Strehl ratio. 
	For the final three test cases, the aim was not to correct all wavefront error present in the system but to leave specific aberrations uncorrected, in a manner which would allow F\&F to solve only LWE-like wavefronts without impacting the system NCPE budget. These cases are discussed in more detail in bullet point 8 below, and in Sect.~\ref{Sec:Disc}.
		
		The following specific observations can be made about the performance of F\&F by comparing the various scenarios presented in Table~\ref{Tab:1_LabResults}:
		\begin{enumerate}
		\setlength\itemsep{3pt}
		\item {\bf Field-of-view:} Shrinking the square FOV available to F\&F from 100 to 32 pixels had only a few percent impact on final PSF quality (rows 9. \& 10.); this was likely due to the removal of high-order wavefront information from the PSF supplied to F\&F.
		\item {\bf Signal-to-noise:} Lowering the image S/N by a factor of 16, from 6380 to 382 (rows 2 \& 7), had a negligible impact on final performance in the high-S/N regime (see Sect.~\ref{Sec:S/N} for a simulated treatment of low-S/N performance).
		\item {\bf Choice of mode basis:} Comparing the use of segmented PTT (row 9.) and 50-mode Zernike (row 8.) bases shows that, as expected, Zernikes were less able to replicate the high-frequency LWE wavefront. There is negligible difference between a PTT-only basis and the full PTT + Zernike basis (row 7.), although the latter is expected to perform better in the presence of additional non-LWE aberrations, or for erroneous pupil rotation angles where the PTT basis is no longer a good description of the LWE wavefront. 
		\item {\bf Using no mode basis:} Comparing the $100\times 100$~pixel FOV PTT basis test (row 10.) with a comparable test applying only Gaussian-filtered (i.e. DM-resolution) F\&F outputs without mode basis projection (row 13.) shows that using the PTT basis actually resulted in a three percent increase in final Strehl ratio. This can be attributed to the effective removal of unwanted high-frequency noise propagation and pupil asymmetry systematics from the F\&F output estimates by the tailored mode basis.
		\item {\bf Wavefront corrector resolution:} Disabling the DM-like Gaussian spatial filtering and applying corrections at the full resolution of the SLM had negligible impact on the final correction performance (rows 10 \& 11). In this case the maximum correctable spatial frequency was effectively set by the FOV of DTTS images. Residual wavefront error maps show high-frequency residuals around the locations of the spiders for filtered wavefront tests as in Fig.~\ref{Fig:4_LabVer}, however these result in only a small amount of additional diffraction along the spiders and hence have a low impact on Strehl ratio.
		\item {\bf Pupil apodisation:} A weak, on-axis pupil apodiser (row 5, and top-centre panel of Fig.~\ref{Fig:3_Apods}) in fact resulted in marginally better performance than the equivalent un-apodised case (row 7.) despite F\&F calculations still assuming a uniform pupil; this is attributed to the suppression of high-order aberrations. F\&F was still stable in the presence of a stronger, asymmetric apodisation of the pupil which even exhibited some vignetting (row 4, and top-right of Fig.~\ref{Fig:3_Apods}) when the algorithm was still not provided with the modified pupil function, showing only a few percent loss in Strehl compared to the above cases. This indicates that F\&F is extremely stable against unknown amplitude aberrations, and even severe pupil mis-alignments which violate the even-pupil assumption implicit in F\&F. For this particular application it is therefore unnecessary to implement additional amplitude retrieval steps (such as the FF-GS extension to F\&F mentioned in Sect.~\ref{Sec:T_FF}) to ensure robust performance in SPHERE.
		\item {\bf Source of phase aberrations:} F\&F was equally capable of correcting strong aberrations from external (non-SLM) sources (row 3.) as it was for SLM-induced aberrations (row 1.), returning both to a final estimated Strehl ratio above 90\%. This indicates a sufficiently accurate orientation, alignment, and phase-to-voltage calibration of SLM commands was achieved for closed-loop correction.
		\item {\bf Applying reference offsets:} Attempting to make F\&F insensitive to PSF centre by removing global tip-tilt components in the output wavefront (row 14.) or attempting to induce specific wavefront reference offsets by manually subtracting them from the F\&F output on each iteration (row 15.) resulted in unstable convergence and were not viable methods for this implementation of F\&F (see below and Sect.~\ref{Sec:Disc} for details). Using the natural lack of sensitivity of a PTT-only basis to low-order Zernike modes to try and correct only the LWE component of an aberrated wavefront (row 16.) resulted in a stable convergence, however the final correction may have included some unwanted partial compensation of non-LWE errors, as the final residual RMS error of 64~nm was smaller than the 77~nm of low-order (i.e. non-LWE) aberrations initially applied on the SLM.
		\end{enumerate}
		As with the main results presented in Fig.~\ref{Fig:4_LabVer}, the various test cases presented in Table~\ref{Tab:1_LabResults} were found to be highly representative of closed-loop simulations directly emulating the conditions of each test case, indicating that F\&F was performing very close to the expected level in this MITHIC implementation. The predictive power of these simulations extends to the identification of the two key limitations so far identified for F\&F; a sensitivity to the centroid zero-point of the PSF in the image (row 14.), and difficulty effectively converging to a non-flat wavefront via direct reference phase map subtraction (row 15.). In both cases divergent behaviour was seen to set in within ten iterations after an initial improvement in wavefront quality, and therefore it is important to evaluate the underlying causes and potential solutions; this analysis is presented in Sect.~\ref{Sec:Disc}.
		
	\subsection{Simulated low-S/N performance of F\&F}
	\label{Sec:S/N}
	
		One of the most important concerns with using the SPHERE DTTS as a focal-plane wavefront sensor is that the low throughput to the DTTS camera results in low S/N images, especially for faint targets. 
		The DTTS control loop is typically operated at a cadence of 1~Hz and is designed to function down to a S/N of approximately ten, and so F\&F should also be stable under these conditions. 
		However, any attempt to sense LWE-like wavefronts from focal-plane images will ultimately be limited by the S/N of the first Airy ring, which corresponds to the dominant spatial frequencies present in this type of wavefront. This forms a significantly stronger S/N constraint than that required for simple tip-tilt correction using the PSF core, and will by necessity limit the efficiency of LWE correction at low S/N.
		
\begin{figure}
		\hspace{-8pt}
		\includegraphics[width=0.5\textwidth]{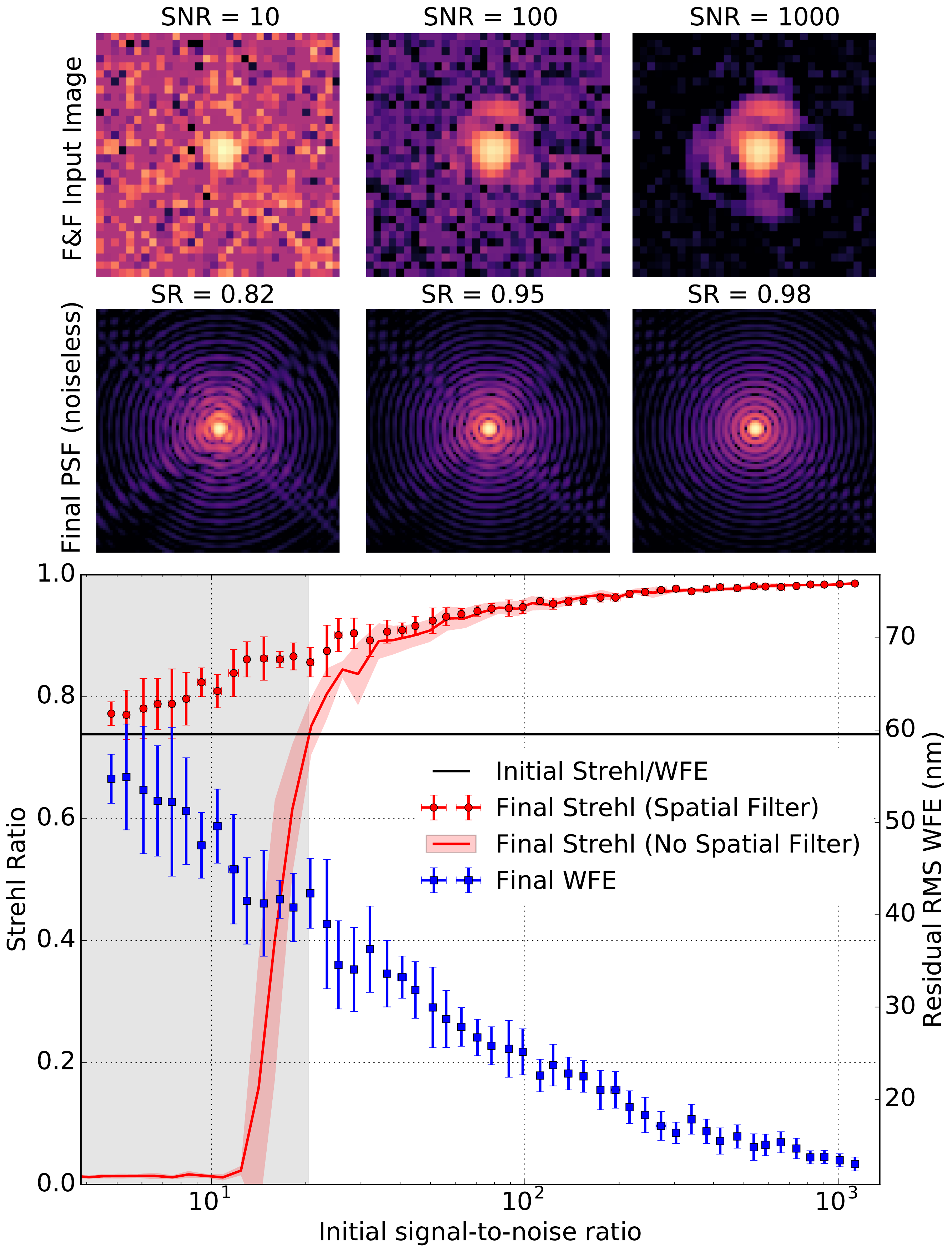}
		\caption{Simulated convergence quality of F\&F as a function of image S/N, for half-wave (319~nm) PVE LWE at a wavelength of 677~nm. {\bf Top image row:} Initial DTTS images at specific S/N, showing increasing visibility of the aberrated first Airy ring. {\bf Bottom image row:} Final noiseless images after 25 F\&F iterations, showing a corresponding improvement in final PSF quality. {\bf Main panel:} Plot of final Strehl ratio (red) and residual RMS wavefront error (blue) as a function of initial S/N. Each point is the average of ten simulations. The black horizontal line denotes both the starting Strehl ratio and wavefront error RMS, of 74\% and 59~nm respectively. The shaded red line shows the Strehl ratio behaviour for F\&F using the full $32~\times~32$ pixel DTTS image as input (i.e. without an adaptive spatial filter), with the shaded region below S/N = 20 denoting the region where this implementation diverges.}
		\label{Fig:5_S/Ncurve}
		\end{figure}

		Figure~\ref{Fig:5_S/Ncurve} illustrates the simulated performance of F\&F as a function of input image S/N, at the laboratory wavelength of 677~nm for comparison with MITHIC results. This was performed in the absence of atmospheric residuals or NCPEs, with only the 319~nm PVE LWE phase aberration (identical to that used for MITHIC bench tests) present. Each data point shows the average final Strehl ratio and wavefront error of ten independent simulations each of 25 iterations, such that the error bars provide an estimate of the post-convergence frame-to-frame stability of F\&F. Here all S/N values are quoted for the central pixel of the PSF, and in all simulations a constant value of $\epsilon = 10^{-3}$ was used for wavefront reconstruction.
		It can be seen from the red (upper) data points that the algorithm is stable over the entire range of S/N values, and still makes some statistical improvement to the wavefront quality even at the minimum S/N of five. Above this the residual wavefront error declines logarithmically, and Strehl ratios of greater than 95\% are achieved for S/N greater than 100. The blue (lower) residual RMS wavefront error data points also provide a useful estimate of the S/N-limited sensitivity of this implementation of F\&F to low-amplitude aberrations. This curve is consistent with the equivalent MITHIC test result (row 2 of Table~\ref{Tab:1_LabResults}), which for an image S/N of 382 achieved 20~nm of residual RMS error. This is equivalent to a sensitivity limit of 45~nm RMS in the H-band, since F\&F operates in radians and so performance can be expected to scale linearly with wavelength.
		
		Such robust performance at low S/N was only possible with the use of an adaptive focal-plane spatial filter to attenuate pixel-to-pixel noise. This filter modifies the input DTTS PSF $p$ by smoothly replacing noise-dominated pixels with the reference (diffraction-limited) PSF $\left|a\right|^2$, such that F\&F sees the higher spatial frequencies as perfect and so provides zero correction on these scales. The resulting filtered PSF $p_{\rm filt}$ is defined by
	\begin{align}
	\label{eq:spatFilt}
	p_{\rm filt} =&~w(r)p  + (1-w(r))\left|a\right|^2~, {\rm where} \\
	w(r) =&~\frac{1}{1+e^{r-r_c}}
	\label{eq:sigmoid}
	\end{align}
	is a radial sigmoid windowing function with a critical cut-off radius of $r_c$. This critical radius is defined to be the point at which the local maxima of the diffraction-limited PSF first drop below a critical S/N of 2.5, which in this instance was found to be an acceptable balance between rejecting noise and preserving wavefront information in the image. It may be possible to further improve the performance of this filter by optimising the functional form for $w(r)$, for instance by using the generalised logistic function. A detailed investigation is however beyond the scope of this paper.
		
	For comparison, the red shaded curve shows the natural behaviour of F\&F in the absence of an appropriate spatial filter, using the full noisy $32~\times~32$ pixel FOV as the input. In this case the pixel-to-pixel noise at high spatial frequencies is directly propagated into strong modal noise in the final correction, resulting in a rapid divergence in wavefront quality below a S/N of 20. With spatial filtering applied it can be seen that the algorithm instead `fails gracefully', simply correcting less of the aberrating wavefront as the corresponding spatial frequencies fall below the image noise threshold. This improved approach also achieves a final post-convergence wavefront quality of better than 40~nm RMS and stability of better than 20~nm RMS $1\sigma$ jitter for all but the lowest S/N values, which can most likely still be improved by more careful optimisation of the adaptive spatial filter profile and cut-off radii as a function of S/N.
		
		For any given observing conditions it should also be possible to further improve F\&F performance by stacking individual 1~s DTTS frames for a longer effective exposure and hence higher S/N, provided that the correction cadence remains significantly shorter than the variability timescale of the LWE. A recent study of the morphology and temporal evolution of the LWE as seen by the SPHERE-IRDIS subsystem (J.F. Sauvage, ESO, private communication, 2017) concluded that under typical conditions (1~m/s wind speed, average LWE of 600~nm PVE, 10~Hz imaging cadence) the majority of the LWE-related structures were coherent on timescales of longer than 10~s, although some small amount of short-term variability was also observed. Frame stacking up to this 10~s threshold would facilitate an additional S/N boost of a factor of three, which even for a faint target with a 1~s exposure image S/N of ten would return image S/N to the regime in Fig.~\ref{Fig:5_S/Ncurve} where we can expect final corrected Strehl ratios of 90\% or greater, corresponding to the removal of the majority of low-frequency LWE error and a significant reduction of spider diffraction effects. In general however, it is advantageous to sense and correct at least ten times faster than the shortest wavefront coherence timescale of interest. This means that F\&F will most likely perform better when operated at the fastest cadence (i.e. with the lowest image S/N) which can be expected from Fig.~\ref{Fig:5_S/Ncurve} to provide sufficiently good correction for any given application.
	
\section{Discussion}
\label{Sec:Disc}

	It is important to discuss the two key limiting factors identified in F\&F, which are found both in simulations and MITHIC lab tests as mentioned in Sect.~\ref{Sec:LabvSim}. The first of these is sensitivity to the centroid location of the image PSF: the tip-tilt zero-point to which F\&F will try to converge is set by the centroid location of the reference PSF $\left|a\right|^2 = \left|\FT[A]\right|^2$, and so the algorithm naturally attempts to apply global tip-tilt corrections if the image centroid differs from this reference. While this tip-tilt correction was observed to be robust up to a dynamic range of approximately 1~radian, it raises the potential for loop conflict if the F\&F zero-point differs from that of the main tip-tilt sensing loop for which the DTTS is primarily used, or if frame-to-frame DTTS tip-tilt correction residuals approach 1~radian. It is therefore desirable to make F\&F completely insensitive to tip-tilt error, for which the na\"{i}ve approach is to directly subtract any measured global tip-tilt components from the output wavefront estimate. However, this approach only results in the build-up of differential tip-tilt between individual VLT pupil segments and a slow divergence of wavefront quality over time. This is because F\&F is not a perfect one-shot phase reconstructor: the exact tilts measured across each pupil segment tend to differ slightly from the global gradient, with the residual between the two still included in the wavefront correction command. As F\&F is still sensitive to the subtracted global tip-tilt error on each subsequent iteration, this residual differential tip-tilt map is re-applied on each iteration and thus builds up steadily over over time. The DTTS image can always be re-centred to pixel precision by shifting the image array (in this case corresponding to a precision of 0.3~$\lD$) to somewhat limit the extent of this centroiding issue, however the best approach is instead to ensure that the reference PSF $p_0$ is constructed to exactly match the zero point of the DTTS to sub-pixel accuracy, thereby ensuring that there are no conflicts between the two correction loops.
	
\begin{figure}
		\includegraphics[width=0.5\textwidth]{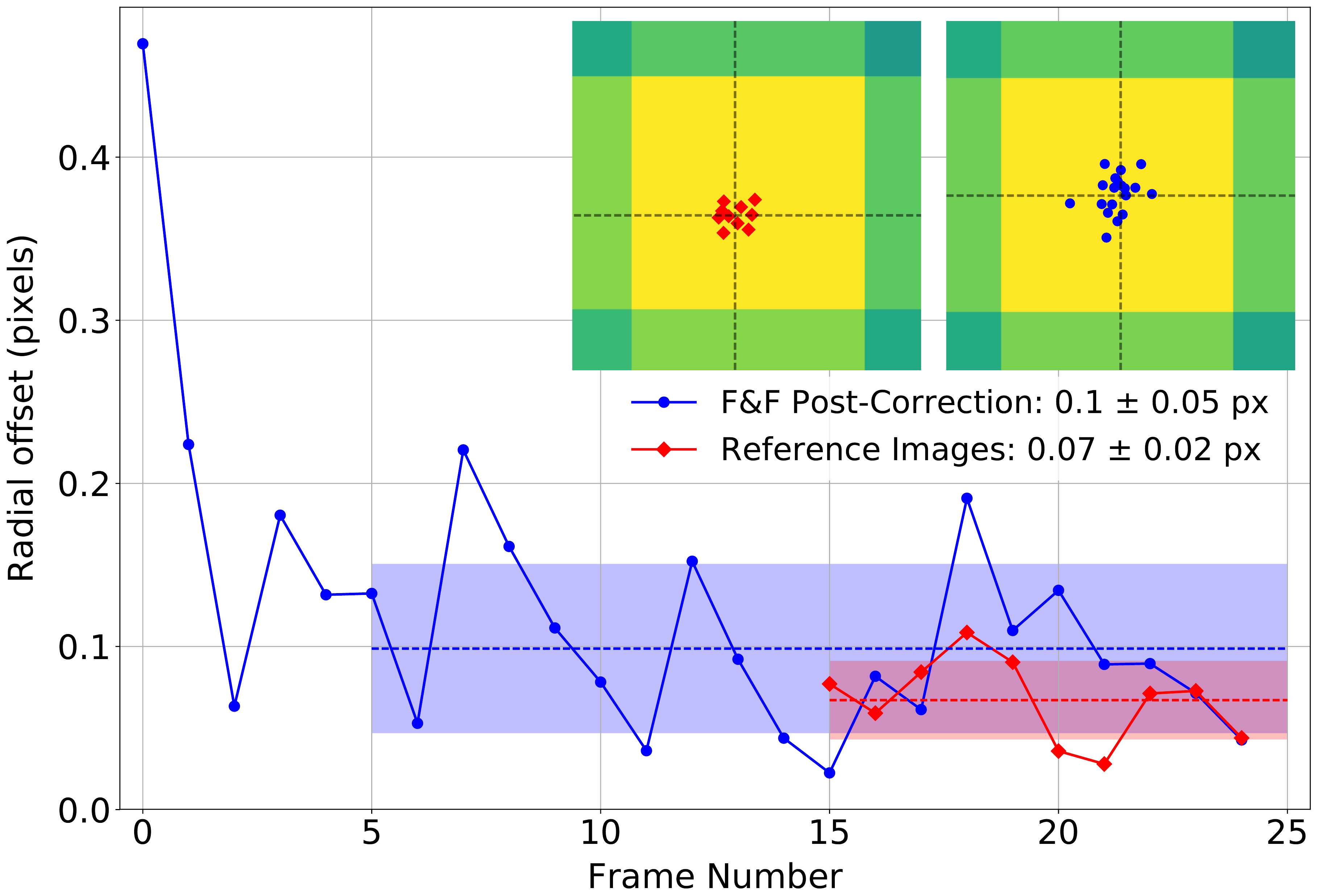}
		\caption{Impact of closed-loop F\&F correction on the tip-tilt stability of the MITHIC bench. {\bf Main panel:} Measured radial offsets from the mean PSF centroid location as a function of iteration number, for the headline MITHIC F\&F convergence test of Fig.~\ref{Fig:4_LabVer} (blue) and a set of reference images containing only natural MITHIC bench image jitter (red). F\&F convergence is taken to be achieved after five iterations. {\bf Inset panels:} The central 0.3~$\lambda$/D pixels of the mean reference image (left) and the F\&F-corrected image (right) respectively, each over-plotted with the centroid data used to compute jitter estimates. Dashed cross-hairs show the mean PSF centroid for each dataset, showing that F\&F applies a systematic sub-pixel offset to align the PSF with its internal reference zero-point.}
		\label{Fig:6_TT}
	\end{figure}

	The other outstanding issue associated with tip-tilt control is whether a stable, converged F\&F control loop adds any additional positional jitter to the PSF, and whether this remains within the specifications of the SPHERE design requirements for coronagraph centring. This was investigated on the MITHIC bench for the main 25-iteration convergence test, previously presented in Fig.~\ref{Fig:4_LabVer} and row 1 in Table~\ref{Tab:1_LabResults}. The absolute deviation of the PSF centroid from its mean location at 3.3~pixels per $\lD$ focal-plane sampling is shown for all frames in Fig.~\ref{Fig:6_TT}, comparing F\&F control with a short reference image sequence at the same cadence, containing only the natural bench image jitter without an active control loop. This shows that F\&F achieved sub-DTTS pixel stability with an RMS of one tenth of a pixel, equivalent to $0.03~\lD$ and hence $1.2$~mas on-sky for SPHERE, only slightly higher the natural bench jitter of $0.02~\lD$ (0.8~mas). 
	It is likely that this result was dominated by normal thermal and mechanical fluctuations in the MITHIC optical path, given the low temporal bandwidth of the manual F\&F control loop and the lack of any other form of active PSF centring control. The on-sky positional stability of SPHERE on the other hand is dominated by the telescope high-frequency vibration environment, with a target of 3~mas (0.07~$\lD$) RMS required for baseline coronagraph operation \citep{Fusco:16}. If it may be assumed that F\&F would add the same amount of additional jitter as was seen in MITHIC, this would constitute only a few percent of the total error budget. Due however to the fact that F\&F would contribute at frequencies between 0.1-10~Hz depending on correction cadence (which is significantly slower than the 10-100~Hz vibrations limiting SPHERE stability), a more detailed investigation would be needed to determine its exact impact on the SPHERE vibrational error budget.
	
	The second known limitation is in attempting to force F\&F to converge to a specific non-flat wavefront, which is in general a useful feature enabling the application of controlled reference wavefront offsets. In this case, the differential optical path between the DTTS and IRDIS focal planes is known to contain 20~nm of static focus error plus some additional higher order NCPEs, which ideally should not be introduced into the science beam by a DTTS-based WFS. While it was found in simulations that the most straightforward approach of subtracting the fixed reference offset from the F\&F output wavefront on each iteration is stable in the noiseless case, under realistic conditions the $\epsilon$ regularisation parameter in Eq.~\ref{eq:v_sol} (which is typically chosen to be comparable to the image noise floor) results in a systematic underestimation of the even wavefront. As for the case of the global tip-tilt drift phenomenon described above, this results in the injection of the residual between the sensed and true offset phase map on each iteration, resulting in the divergent behaviour seen in row 14 of Table~\ref{Tab:1_LabResults}. While it is possible to systematically ignore specific modes entirely (for example focus or astigmatism) by subtracting the measured coefficient of that mode from the raw F\&F output wavefront before final mode basis projection, it is unclear how this would affect the efficiency of LWE correction: the segmented PTT basis is not fully orthogonal to such Zernike modes, and astigmatism in particular is typically present in the LWE wavefronts we wish to correct. 
	
	Attempts to modify the F\&F algorithm itself to properly treat reference offsets are ongoing, however this is made challenging by the fact that the simplifying assumption of a real, even pupil function necessary for an analytical solution also prevents direct modification of the target PSF to include reference phase aberrations, i.e. $a = \FT[A] \rightarrow \FT[A{\rm e}^{i\phi_{ref}}]$. This is because a number of terms in the original Taylor expansion of the PSF which correspond to non-even and complex aperture terms are deliberately neglected from the derivation before arriving at Eq.~\ref{eq:PSF_approx} in order to obtain an analytical solution for the focal plane fields. Because these neglected terms now become significant, such a substitution for $a$ is no longer valid. A generalised version of F\&F capable of arbitrary wavefront reference offsets would also be of great interest for operation with coronagraphic images, especially the apodising phase plate (APP) coronagraph \citep{Kenworthy:10,Snik:12,Otten:17}. However, given the low level of NCPEs in the SPHERE case of interest there would be a minimal impact on science image Strehl ratio associated with allowing free convergence to the DTTS focal-plane: the small degradation in image quality is expected to be vastly outweighed by the gain in raw contrast performance from controlling the LWE. In high-S/N environments, it would also be expected that using F\&F to stabilise NCPEs up to the NIR coronagraph with a general Zernike mode basis would also provide an improvement in final high-contrast imaging performance despite inducing 20~nm of static focus error into the science beam.
	
	The results presented in this paper are somewhat idealistic in that they assume the XAO system is composed of the sole DM component to which the output phase commands from F\&F can be accurately implemented via phase conjugation. In reality, correction must be achieved by modifying reference slope offsets on the SAXO SH-WFS during operation of the main XAO sensing and correction loop. In addition to the potential for control loop conflicts, it is currently unknown how this approach will filter the high spatial frequencies present in the LWE wavefront, and at what point the finite dynamic range of the SH-WFS will limit the correction of high-amplitude LWE cases. The remaining stroke on the SAXO DM during closed-loop operation will also determine how effectively the highest amplitude LWE cases can be corrected. For example, an 800~nm PVE LWE would constitute 11\% of the total $\pm3.5~\mu\rm{m}$ SPHERE DM stroke and 26\% of the $\pm1.2~\mu\rm{m}$ inter-actuator stroke \citep{Fusco:06}. Since the LWE occurs under good seeing conditions where there is less strain on the AO system, it is likely that almost all LWE cases would see significant improvement before being limited by DM saturation. 
	Due to the complicating factors listed here, applying any form of focal-plane wavefront control using the SPHERE DM and SH-WFS is clearly still an important area to be addressed, and may require the development of a dedicated control scheme. However, initial tests have been encouraging in showing that the DM can accurately reconstruct a strong differential piston via the reference offset approach in both open and closed loop, with the width of the phase discontinuity boundary at pupil segment edges consistent with the influence function of the DM actuators \citep{Sauvage:16}. 
	
	Altogether it is expected that with appropriate calibration the current implementation of F\&F is capable of providing at least an order of magnitude of raw contrast improvement in SPHERE coronagraphic imaging performance at 2-4$\lD$ in typical LWE-affected cases for S/N greater than 20, by returning the distorted first Airy ring to near-diffraction-limited performance (see Fig.~\ref{Fig:2_CoroEff}). Additional gains may also be made in post-processing if F\&F has sufficient image S/N to stabilise quasi-static speckle structure, allowing for more effective reference PSF subtraction or removal via ADI or PCA-based PSF subtraction techniques.
	By running in continuous closed-loop mode, on-sky performance can also be expected to be superior to that presented in Fig.~\ref{Fig:5_S/Ncurve} and \cite{Wilby:16a}, since both this work and the previous study are concerned with compensating an established LWE wavefront error. Provided that F\&F is operated above the critical cadence of ten times the variability timescale of PTT wavefront errors as they arise. In low-S/N cases a piston-only correction loop can still be expected to reduce the impact of many LWE wavefronts on the coronagraphic PSF, and is also less likely to conflict with the main AO loop as the SH-WFS is in principle insensitive to differential piston aberrations.

\section{Conclusions}
\label{Sec:Conc}

	We have demonstrated that the Fast \& Furious sequential phase diversity algorithm is capable of robustly eliminating strong LWE wavefronts in the MITHIC high-contrast laboratory testbench environment, where it reliably returned image Strehl ratios to better than 90\% within five closed-loop iterations. This was achieved in the presence of strong static low-order aberrations, low S/N, and small FOV images representative of the SPHERE DTTS, but in the absence of incoherent atmospheric speckle residuals or an active primary XAO loop, and assumed an idealised SPHERE DM response for correction.
	We find no significant discrepancies between these MITHIC laboratory results and the predictions of dedicated LWE simulation code \citep{Wilby:16a} designed to emulate focal-plane wavefront sensing with the DTTS sensor. Therefore, further work carried out using this code is expected to be representative of performance achievable with F\&F on-sky with the SPHERE instrument.
	
	Supporting simulations showed that this implementation of F\&F is also stable over the full working S/N range of the DTTS sensor down to at least S/N~=~5, and is capable of efficiently removing the dominant structures of the LWE for ${\rm S/N \geq 20}$. For targets where this condition can be satisfied for correction cadences faster than the dominant LWE variability timescale (estimated from IRDIS observations to be longer than 10~s) an on-sky implementation of F\&F should be capable of effectively maintaining a near-diffraction-limited PSF core under the strongest LWE conditions routinely seen by SPHERE.
	Such an improvement is expected to provide at least an order of magnitude gain in raw contrast over typical LWE-affected PSFs close to the coronagraphic inner-working angle, greatly improving the ultimate contrast performance of the SPHERE instrument under the best seeing conditions. 
	
	Further efforts will focus on understanding the interplay between F\&F and a realistic AO environment, including the spatial filtering properties of WFS reference slope offset based control, and the potential for conflicts in a multi-control loop system. It is also of great interest to develop a generalised version of the algorithm which is capable of converging to arbitrary non-flat reference wavefronts: in addition to providing greater flexibility for closed-loop control, this would also allow F\&F to operate directly with many types of coronagraphic science image.
	 
	In addition to this specific application for controlling the LWE in SPHERE, the stability and versatility of F\&F makes the algorithm highly suitable for other real-time focal-plane wavefront control tasks, such as NCPE correction or mirror co-phasing of segmented telescopes, provided that narrowband image data is available at a sufficiently fast cadence.
	With instances of the more general `island effect' now being repeatedly seen in high-contrast instruments beyond the VLT, it is clear that focal-plane wavefront control methods such as F\&F could become increasingly essential for the field. This will be especially important for the upcoming ELT-class telescopes, which will feature highly segmented pupils and a large amount of obscuring support structure. It can be expected that these telescopes will be more prone to island effect and LWE phenomena than simple four-quadrant pupil geometries, and without appropriate mitigation strategies these effects may severely limit the performance of their XAO-fed high-contrast instruments. The computational simplicity of F\&F allows it to scale efficiently to work with high-resolution deformable elements, and in principle makes it sufficiently fast for high-speed (kHz) wavefront control applications: this makes it an attractive focal-plane phase control solution for both current and future instruments.

\begin{acknowledgements}
The authors would like to thank the anonymous referee, whose helpful comments have significantly improved the contents of this paper.
This work is funded by the Nederlandse Onderzoekschool Voor Astronomie (NOVA), via the NOVA-4 EPICS R\&D support grant, and is also supported by a grant from the French Labex OSUG@2020 (Investissements d’avenir – ANR10 LABX56).
SPHERE is an instrument designed and built by a consortium consisting of IPAG (Grenoble, France), MPIA (Heidelberg, Germany), LAM (Marseille, France), LESIA (Paris, France), Laboratoire Lagrange (Nice, France), INAF - Osservatorio di Padova (Italy), Observatoire astronomique de l’Université de Genève (Switzerland), ETH Zurich (Switzerland), NOVA (Netherlands), ONERA (France) and ASTRON (Netherlands) in collaboration with ESO. SPHERE was funded by ESO, with additional contributions from CNRS (France), MPIA (Germany), INAF (Italy), FINES (Switzerland), and NOVA (Netherlands). SPHERE also received funding from the European Commission Sixth and Seventh Framework Programmes as part of the Optical Infrared Coordination Network for Astronomy (OPTICON) under grant number RII3-Ct-2004-001566 for FP6 (2004-2008), grant number 226604 for FP7 (2009-2012), and grant number 312430 for FP7 (2013-2016). 
\end{acknowledgements}

\bibliographystyle{aa}
\bibliography{Wilby18_LWE}

\end{document}